\documentclass[10pt,journal,compsoc]{IEEEtran}

\ifCLASSOPTIONcompsoc
  \usepackage[nocompress]{cite}
\else
  \usepackage{cite}
\fi

\ifCLASSINFOpdf
   \usepackage[pdftex]{graphicx}
   \DeclareGraphicsExtensions{.pdf,.jpeg,.png}
\else
   \usepackage[dvips]{graphicx}
   \DeclareGraphicsExtensions{.eps}
\fi

\usepackage{amssymb,amsthm,bbm}
\usepackage[cmex10]{amsmath}

\ifCLASSOPTIONcompsoc
  \usepackage[caption=false,font=footnotesize,labelfont=sf,textfont=sf]{subfig}
\else
  \usepackage[caption=false,font=footnotesize]{subfig}
\fi

\usepackage{url}

\usepackage{booktabs} 
\newcommand{\cC}{{\mathcal{C}}}

\newtheorem{thm}{Theorem}[section] 
\newtheorem{definition}[thm]{Definition}

\makeindex
\begin{document}
\title{Measuring Influence on Instagram: a Network-oblivious Approach}

\author{Noam~Segev, Noam~Avigdor, Eytan~Avigdor
\IEEEcompsocitemizethanks{
\IEEEcompsocthanksitem N. Segev, N. Avigdor and E. Avigdor are with Klear Influencer Marketing\protect\\
E-mail: [noam.segev, n, e]@klear.com}
}

\markboth{Measuring Influence on Instagram: a Network-oblivious Approach}%
{Measuring Influence on Instagram: a Network-oblivious Approach}

\IEEEtitleabstractindextext{%
\begin{abstract}
This paper focuses on the problem of scoring and ranking influential users of Instagram, 
a visual content sharing online social network (OSN). 
Instagram is the second largest OSN in the world with 700 million active Instagram accounts, 
32\% of all worldwide Internet users\footnote{\url{https://www.omnicoreagency.com/instagram-statistics}}.
Among the millions of users, 
photos shared by more influential users are viewed by more users than posts shared by less influential counterparts. 
This raises the question of how to identify those influential Instagram users.

In our work, we present and discuss the lack of relevant tools and insufficient metrics for influence measurement, 
focusing on a network oblivious approach and show that the 
graph-based approach used in other OSNs is a poor fit for Instagram. 
In our study, we consider user statistics, some of which are more intuitive than others, 
and several regression models to measure users' influence.
\end{abstract}

}

\maketitle

\IEEEraisesectionheading{\section{Introduction}\label{sec:intro}}


\IEEEPARstart{T}{he} he transition to Web 2.0 transformed the business models of online marketing 
from a global ad approach based to individual opinions and targeted campaigns \cite{morgan1994commitment, scoble2006naked, Abdullah:Incorporating, gegenhuber2017making}. 
Web 2.0 not only took traditional marketing strategies to the extreme via viral marketing campaigns \cite{kurultay2012dynamics, wilde2013viral, rakic2014viral}, 
but it also gave rise to new techniques of brand building and audience targeting via influencer marketing \cite{brown2013influence, zietek2016influencer}. 
In fact, the use of micro-influencers, trusted individuals within their communities, 
has been seen as a more effective way to build a brand in terms of audience reception and return on investment 
\cite{ha2015experiment, bijen2017ad, lisichkova2017impact}.

Instagram, which is a visual content sharing online social network (OSN), has become a focal point for influencer marketing.
With power users and micro-influencers publishing sponsored content companies need to rate these influencers and determine their value\cite{richardson2013cigar, cornet2016instagram, chen2016rise}.
Most of today's scoring themes rely on graph-based algorithms of a known network graph. 
Such graphs are not always available, and building them for Instagram users requires a great deal of resources, 
e.g., crawling time and computing costs.
A possible solution would be to infer the underlying network structure using the user activity logs, 
as described by Barbieri et al.\cite{barbieri2013influence}, 
but even in the event a graph is constructed it would not necessarily be of much use 
given that information decays exponentially along the graph even under optimal passive information propagation, 
which is not the case.

The rest of the paper is organized as follows: 
In Section~\ref{sec:background} we described OSNs in greater detail as well as current influence measuring schemes. 
We then present our annotations and formal description of the problem of measuring and ranking influence in Section~\ref{sec:prob}.
The dataset of Instagram users and their posts is described in Section~\ref{sec:Dataset}, 
followed by discussion on the extracted and aggregated features of the testable data in Section~\ref{sec:Features}.
Following this, we present our testing methodology, baselines, regression models and experimental results in Section~\ref{sec:EXPERIMENTAL}.
Finally, we discuss our conclusion and possible future work in Section~\ref{sec:conclusions}.

\section{Background and Related Work}
\label{sec:background}
Online social media networks are often described as a directed graph 
with entities such as users acting as nodes and relationships as the edges.
Such edges can be unidirectional or bi-directional, e.g., 
an Instagram "follower" and a Facebook "friendship", respectively.
These edges do not need to represent a long-lasting relationship; 
they can signal a one-time engagement, e.g., a "like" or a "comment".
Following this, link prediction in OSNs became an active research field 
focused on community detection, in the case of users as nodes \cite{Budalakoti:2012:BIF:2187836.2187932,Hsieh:2013:OOS:2488388.2488439}, 
or content suggestion otherwise \cite{Amin:2012:SRL:2365952.2366013, Reda:2012:MSR:2396761.2396847, Rodriguez:2012:MOO:2365952.2365961}.

In most OSNs, user-generated content is ``pushed", i.e., propagated via interaction.
When a user uploads a post, their followers can see the post and choose to pass it along, 
creating a pyramid-formed cascade of information. 
Thus, if user A follows user B who, in turn, follows user C, and user C posts some content user B chooses to share, 
user A is passively influenced by user C.
These social micro-networks tend to grow around influential, active users \cite{ethan2009differing, lu2016vital}. 
Instagram content, however, is ``pulled", i.e., information propagation requires activity along the pyramid, 
such that, using our earlier example, for user C's post to reach user A, user A must look for content suggested by trusted users.

This situation raises the question of how to rank users in OSNs.
As OSNs are traditionally described as graphs, ranking has been done using various graph statistics, 
from simple in/out degree to node closeness \cite{cha2010measuring, chen2012identifying}, 
as is the case with the work of Anger and Kittl in Twitter \cite{anger2011measuring} 
and Agarwal et al. in the context of influential blogs \cite{agarwal2008identifying}.
Other techniques extend to existing link analysis algorithms - 
the most popular one being PageRank \cite{Bianchini:2005:IP:1052934.1052938, Brin:1998:ALH:297805.297827}.
Weng et al. suggested twitterRank \cite{weng2010twitterrank} and Khrabrov et al. introduced starRank \cite{khrabrov2010discovering}, 
both extensions of PageRank working on Twitter's follower and engagements graphs, respectively.
On LinkedIn, a professional OSN, 
Budalakoti and Bekkerman suggested a fair-bets model for ranking via transfer of authority \cite{budalakoti2012bimodal}, 
and on Instagram Egger suggested a PageRank extension for influencer ranking \cite{egger2016identifying}.

On Instagram, unlike Twitter, follower data is not publicly available. 
While this information can still be collected via crawling, it is a long and expensive process. 
A possible solution would be to infer the underlying network structure using the user activity logs, 
as described by Barbieri et al.\cite{barbieri2013influence}. 
Even in the event a graph is constructed, it would not necessarily be of much use 
given that information decays exponentially along the graph even under optimal passive information propagation, 
which is not the case.



\section{Problem Formulation}
\label{sec:prob}
The influence of a user in an OSN has been described either in simple, intuitive measures
or as a non-intuitive measurable graph statistic with no real-world meaning \cite{weng2010twitterrank, goyal2010learning, egger2016identifying}.
One such measure is the user's expected post engagements. 
We extend this definition in the realization that being exposed to specific content often does not lead to active engagement.

We say the influence of an Instagram user (\textit{Instagrammer}) is the expected exposure their content would receives, 
or, their expected number of views per post. 
Adhering to the law of large numbers, we can estimate the users' influence using Definition~\ref{def:influence}:
\begin{definition}
\label{def:influence}
 Let $U$ be the set of all Instagrammers, $\cC$ be all content posted on Instagram, 
 $v_c$ the number of Instagrammers that saw post $c \in \cC$ and $\cC_u \subset \cC$ is the content posted by $u \in U$.
 We say that the \textit{influence} of Instagrammer $u$ is:
 \begin{displaymath}
   Inf_u = \frac{\sum_{c \in \cC_u} v_c}{\left| \cC_u \right|}.
 \end{displaymath}
\end{definition}


\section{Instagram Dataset}
\label{sec:Dataset}

For the purpose of this study, a set of Instagram data was prepared in April 2017, 
including posts published during 2015-2016 but prior to September 2016.
We focused on a subset of Instagram posts where view counts were accessible.
Independent studies have shown that 50\% of engagements of an Instagram post happen within 72 minutes of publication and 95\% within the following week\footnote{\url{https://blog.takumi.com/the-half-life-of-instagram-posts-3db61fb1db75}}. 
As the change of feed ranking in March 2016 did not cause statistically significant changes to activity, 
and as all posts examined by us were over 6 months old, 
we say that the data is stable, meaning, all posts have reached at least 95\% of their potential views and engagements.
The data was prepared as follows:
\begin{enumerate}
 \item We gathered information on videos 
  \footnote{We used Instagram API to collect user statistics. 
   We did not use the API to gather data for the posts themselves due to API limits.
   Instead, we parsed each post web-page.}
  published by a set of randomly selected Instagrammers with publicly accessible profiles. Denote the set of users as $U$.
  Each of these Instagrammers must have published a minimum of 10 video posts before September 2016.
 \item For each video $c \in \cC$, we collected the following metrics:
  \begin{itemize}
   \item \textit{$likes_c$} - Number of likes awarded to post $c$.
   \item \textit{$comments_c$} - Number of comments given to post $c$.
   \item \textit{$v_c$} - Number of Instagrammers who watched part of the video.
  \end{itemize}
\end{enumerate}
A total of $940,439$ posts by $115,044$ Instagrammers was collected\footnote{
This collection of anonymized public information is available at 
\url{https://klear.com/sigir/instagram_data.zip}}.

\subsection{Instagram Statistics}

\begin{figure*}
 \centering
 \subfloat[][\centering Views Histogram\label{fig:distViews}]{
  \includegraphics[width=0.24\textwidth]{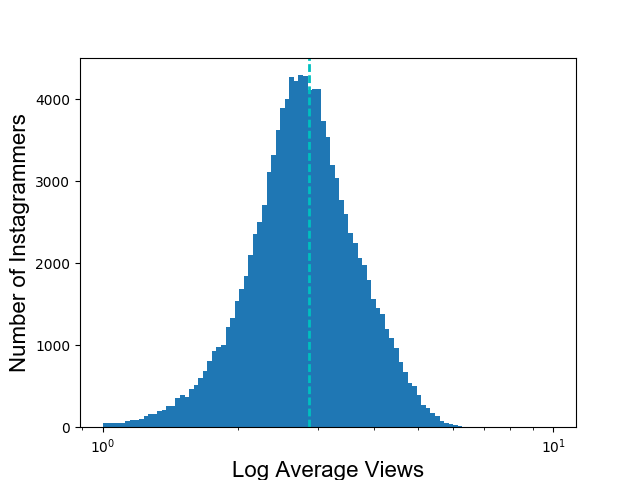}
 }
 \subfloat[][\centering Views per Followers\label{fig:distFollowers}]{
  \includegraphics[width=0.37\textwidth]{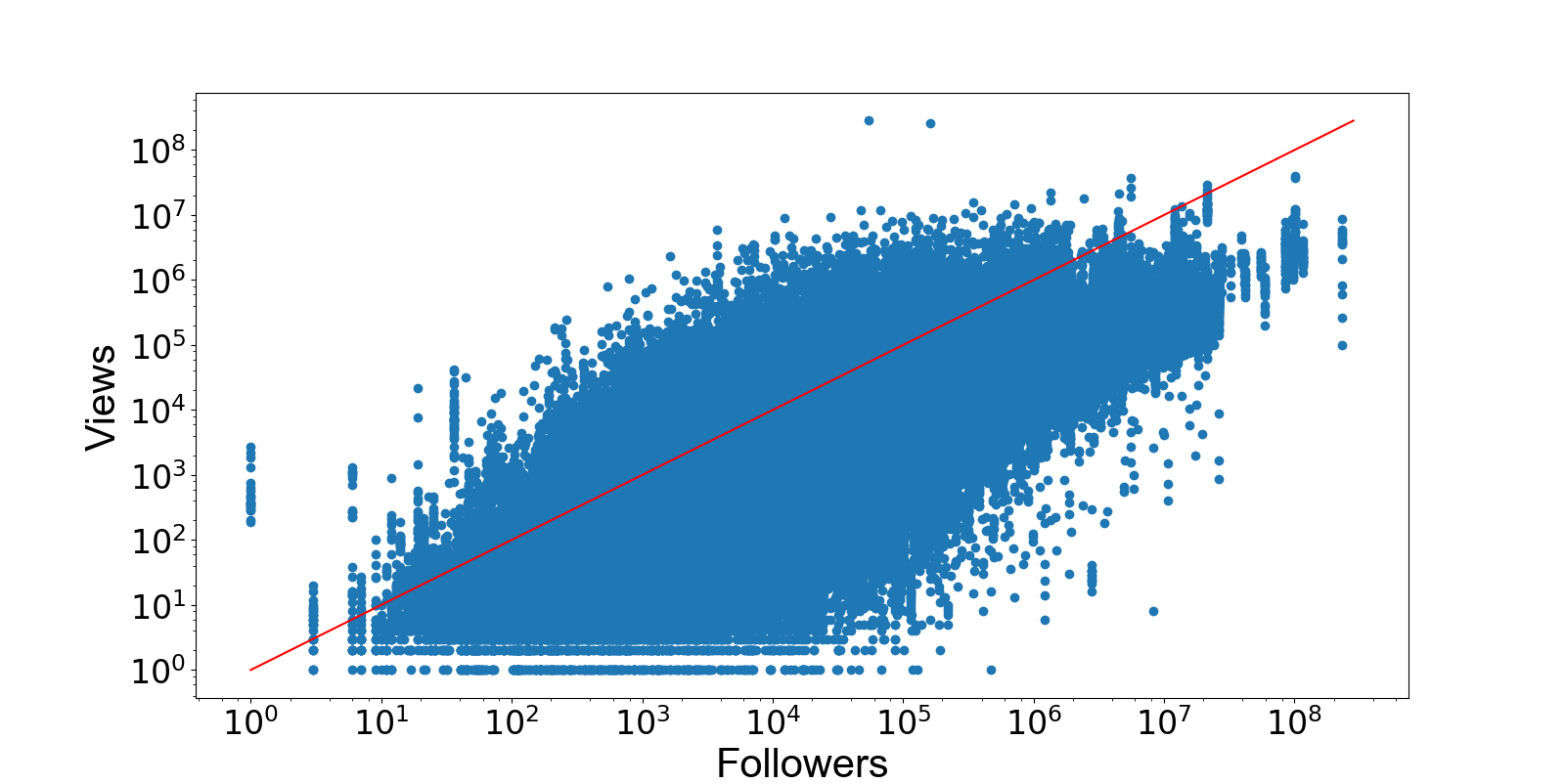}
 }
 \subfloat[][\centering Views per Likes\label{fig:distLikes}]{
  \includegraphics[width=0.37\textwidth]{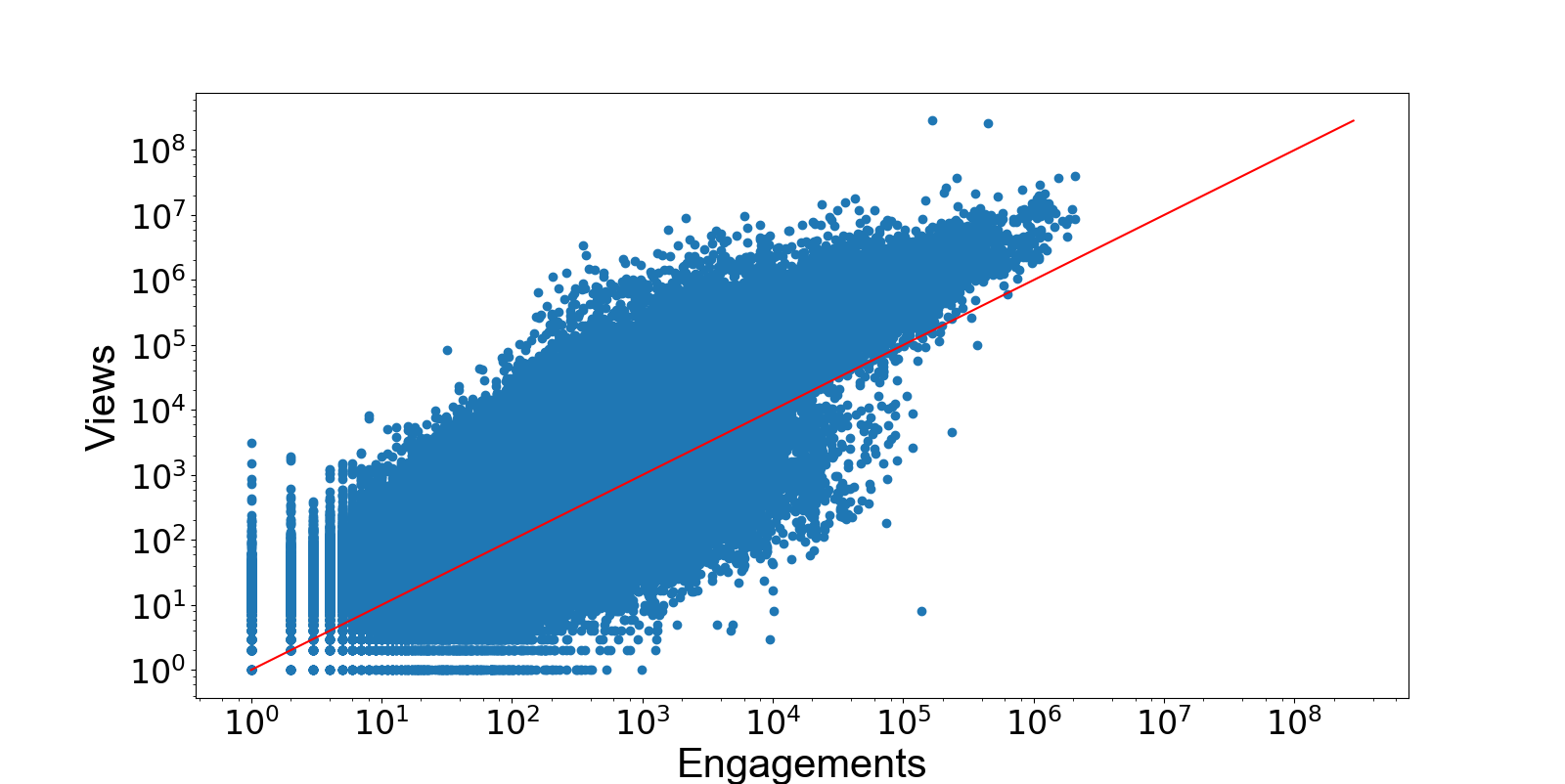}
 }
 \caption{Distributions per Instagrammer}
\end{figure*}

The distribution for log average views per Instagrammer is presented in Figure~\ref{fig:distViews}, 
from which we can tell that this statistic behaves in a log-normal distribution with a mean of $748$ views. 
Furthermore, as this distribution is so close to normal, 
we ascertain that our selection of sampled Instagrammers is a good semblance of real-world influence 
with micro-influencers populating the dense mean 
and casual users and celebrities appearing at the distribution extremes.

Post views per followers and per engagement appear in Figures~\ref{fig:distFollowers} and \ref{fig:distLikes}, respectively; 
these show some underline truths of Instagram. 
It can be seen that normally, the number of followers a user has outnumber his views, 
as we expect following the described flow of information. 
However, we found that this is not the case for sponsored posts, 
massively engaged content or externally referenced content.
Another unlikely situation is of posts having more engagements than views.
This relates either to bought engagements, often via automation tools and fake accounts, 
or to an interesting phenomenon on Instagram known as 
"Like You, Like Me" where content is engaged simply to reciprocate prior engagements.
The issue mitigates as the number of engagements increase.

To avoid these sorts of odd behaviors, we performed univariate outliers removal, 
ignoring the top and bottom posts for users with posts statistics above 2 standard deviations.

\subsection{Features Collected}
\label{sec:Features}
For the purpose of this work, we collected basic features directly from Instagram. 
Expanding on the posts features mentioned above, we also collected user specific statistics. 
We then considered each user as a data point with the following statistics:
\begin{itemize}
 \item \textit{$likes$} - The average number of user post likes.
 \item \textit{$comments$} - The average number of comments per user post.
 \item \textit{$followers$} - The users audience size.
 \item \textit{$\sqrt{likes \cdot followers}$} - Geometric mean of likes and followers, 
   taken as neither statistic is an exact representation of influence.
 \item \textit{$\frac{followers}{post}$} - Used to suggest odd behavior 
   as same level influencers should have similar ratios.
 \item \textit{$\frac{comments}{likes}$} - Another odd behavior indicator
   as bought engagements tend to effect likes more than comments.
 \item \textit{$focus$} - The difference and ratio between most and least engaged post, 
   these features were designed to test the variance and stability of a user engagement level.
\end{itemize}

\section{Regression Models}
We attempt to measure influence using well known regression models 
via the features described at Section~\ref{sec:Features}. 
Furthermore, as some models are sensitive to redundant features, 
we perform recursive feature elimination, generating a subset of informative features
for the problem at hand. 

The models tested include:
\begin{itemize}
 \item \textit{Ridge Regression(RR)} - An extension of Linear Regression, 
   RR attempts to overcome Linear Regressions' problem with feature multi-collinearity 
   adding l2 norm regularization of the coefficients to the minimization problem\cite{hoerl1970ridge}.
 \item \textit{Random Forest(RF)} - Non-linear algorithms that rely on 
   ensembles of decision trees with randomness injected into the model in both features 
   and instances selection\cite{breiman2001random}.
\end{itemize}

We also introduce a meta-algorithm expansion of our own. 
It is clear that not all influencers should be handled in the same manner 
and celebrities statistics would show vast differences than those of micro-influencers.
We propose a Multiple-Regression model, where data is separated to subsets, 
in our case, using the K-Means clustering algorithm on the followers statistic \cite{forgy1965cluster}, 
and building a regression model for each subset.

Finally, it can be seen in Figures~\ref{fig:distFollowers} and \ref{fig:distLikes} 
that the likes and followers' statistics grow in an exponential manner. 
To handle potential bias towards these features, both in clustering and regression, 
we transform these statistics using a log scale, i.e., $f\left(x\right)=\frac{x}{\ln x}$.

\section{Experimental Results}
\label{sec:EXPERIMENTAL}

In this section, we present the methodology for evaluating different techniques 
and introduce two simple yet commonly used baselines. 
We test our models and present the results of our attempt to measure the influence of Instagram users. 

\subsection{Methodology}

To compare between different models, we employ two commonly used statistics. 
To test the model's ability to measure influence, we employ the coefficient of determination, denoted $R^2$. 
Bound by 1, higher $R^2$ scores would indicate lower error variance which indicates a tighter model.
Comparing the order of the predicted influence with the real influence allows us to rank users.
To test the resulting ranking created we use Spearman's rank correlation coefficient, denoted $r_s$. 

To avoid the problems of a model tuned specifically to the test data, 
we use a five-fold cross validation technique.
We randomly split $U$ into five equally sized sets of disjoint Instagrammers 
and use them as five train-test datasets, 
each test set contains roughly 20\% the size of the original set of users $U$ 
and the train set is made of the remaining 80\%.
The results are averaged on the five test cases.

\subsection{Baselines}
Two natural baselines for measuring influence are to use 
the user's audience size (followers) or engagement level (number of likes).
We use both statistics baselines, utilizing a Linear Regression model.

While outside our scope, 
for completeness purposes, we used the PageRank extension suggested by Egger \cite{egger2016identifying}.
For this, we crawled Instagram, creating a commentators graph around our test users.

\subsection{Comparison of Techniques}
The results of the $R^2$ and $r_s$ statistics for the regression models 
and baselines are provided in Table~\ref{tab:regression}. 
These results include both clustered and unclustered attempts, 
as well as, show the result of the feature reduced models.

It is clear that the followers statistic, while intuitive and is often used in real-world scenarios, 
is the weakest on any given metric. 
This correlates with previous findings by Cha et al.\cite{cha2010measuring}.
The engagement baseline is the best choice for a direct ranking approach as
it is almost the best, certainly within error range, 
and is much simpler to use than the full regression models.

Amongst our suggested models, Multi-Regression was not a useful approach 
while feature reduction still resulted in strong models with only half the features.
When comparing RR and RF, we clearly see that RR is a more accurate model. 
This is due to a limitation of the RF model - while RR can return any possible value, 
RF models can return only linear combination of values in the training set 
and while this result in a better ranker, the predicted value more often overshoots.

Due to resource and time constraints we ran the PageRank algorithm a subset of 10\% of the users, 
resulting in an $r_s$ score of 0.673.
These results, only better then the followers baseline, are to be expected given Instagram's flow of information, 
as discussed in Section~\ref{sec:background}.

\begin{table} [bth]
 \caption{$R^2$  and $r_s$ statistics for regression models}
 \label{tab:regression}
 \begin{tabular}{l|c|c|c|c}
  & \multicolumn{2}{|c|}{Regression} & \multicolumn{2}{|c}{Multi-Regression} \\ \hline
  & $R^2$ & $r_s$ & $R^2$ & $r_s$ \\ \hline
  full Ridge Regression & $\bf{0.725}$ & $0.848$ & $\bf{0.727}$ & $0.821$ \\ \hline
  full Random Forest & $0.626$ & $\bf{0.869}$ & $0.621$ & $\bf{0.861}$ \\ \hline
  minimal Ridge Regression & $0.723$ & $0.818$ & $0.727$ & $0.818$ \\ \hline
  minimal Random Forest & $0.616$ & $0.864$ & $0.611$ & $0.859$ \\ \hline
  \hline
  Followers Baseline & $0.211$ & $0.757$ & $0.204$ & $0.725$ \\ \hline
  Likes Baseline & $0.666$ & $0.859$ & $0.654$ & $0.853$
 \end{tabular}
\end{table}

\section{Conclusions and Future work}
\label{sec:conclusions}

This work focused on measuring influence and influencer ranking on Instagram, a content sharing OSN.
Our definition of influence (Def.~\ref{def:influence}) and the features extracted from public information 
allowed us to use out-of-the-box regression models to create what is, to our knowledge, 
the first influence ranking algorithm based on an intuitive score derived from network-oblivious statistics.
We have shown general truths regarding Instagram such that the commonly sought out audience size is a poor metric for influence.

In our work, we did not consider the temporal nature of influence, i.e., 
the influence of a user is likely to change over time. 
The rate of change may even depend on the influence itself, 
as per the rich get richer phenomenon \cite{araujo2014not}.

Lastly, only simple user and posts statistics were used in this work. 
We believe the use of more complex features would result in stronger models and a better ranking algorithm.
These features can be post specific, from the simple "day of the week" to complex "contains faces" \cite{silva2013picture, bakhshi2014faces}, 
user specific, e.g. the user's age or common content type \cite{jang2015generation, hu2014we}, 
or features relating to a user's audience, such as audience location or age \cite{ferrara2014online,manikonda2014analyzing}.




\bibliographystyle{IEEEtran}
\bibliography{instagram} 

\end{document}